\documentclass[twocolumn,showpacs,prl,preprintnumbers, amsmath,amssymb,superscriptaddress]{revtex4}

 \usepackage{graphicx}
 \usepackage{amsmath}
 \usepackage{amsfonts}
 \usepackage{amssymb}
 \usepackage{pstricks}
 \usepackage{psfrag}
 \usepackage{epsfig}
 \usepackage[ansinew]{inputenc}

\begin{document}

\title{Amoeboid motion in confined geometry}


\author{Hao Wu}
\thanks{These authors equally contributed to this work}
\affiliation{Universit{\'e} Grenoble Alpes, LIPHY, F-38000 Grenoble, France}
\affiliation{CNRS, LIPHY, F-38000 Grenoble, France}
\author{M. Thi\'ebaud}
\thanks{These authors equally contributed to this work}
\affiliation{Universit{\'e} Grenoble Alpes, LIPHY, F-38000 Grenoble, France}
\affiliation{CNRS, LIPHY, F-38000 Grenoble, France}
\author{W.-F. Hu}
\affiliation{Department of Applied Mathematics, National Chiao Tung University, 1001 Ta Hsueh Road, Hsinchu 300, Taiwan}
\author{A. Farutin}
\affiliation{Universit{\'e} Grenoble Alpes, LIPHY, F-38000 Grenoble, France}
\affiliation{CNRS, LIPHY, F-38000 Grenoble, France}
\author{S. Rafa\"i}
\email{salima.rafai@ujf-grenoble.fr}
\affiliation{Universit{\'e} Grenoble Alpes, LIPHY, F-38000 Grenoble, France}
\affiliation{CNRS, LIPHY, F-38000 Grenoble, France}
\author{M.-C. Lai}
\affiliation{Department of Applied Mathematics, National Chiao Tung University, 1001 Ta Hsueh Road, Hsinchu 300, Taiwan}
\author{P. Peyla}
\affiliation{Universit{\'e} Grenoble Alpes, LIPHY, F-38000 Grenoble, France}
\affiliation{CNRS, LIPHY, F-38000 Grenoble, France}
\author{C. Misbah}
\affiliation{Universit{\'e} Grenoble Alpes, LIPHY, F-38000 Grenoble, France}
\affiliation{CNRS, LIPHY, F-38000 Grenoble, France}


\begin{abstract}
Many eukaryotic cells undergo frequent shape changes (described as amoeboid motion) that enable them to move forward. We investigate the effect of confinement on a minimal model of amoeboid swimmer. Complex pictures emerge: (i) The swimmer's nature (i.e., either pusher or puller) can be modified by confinement, thus suggesting that this is not an intrinsic property of the swimmer. This swimming nature   transition stems from intricate  internal degrees of freedom of membrane deformation.  (ii) The swimming speed might increase with increasing confinement before decreasing again for stronger confinements. (iii)  A  straight amoeoboid swimmer's trajectory  in the channel can become unstable, and ample lateral excursions of the swimmer prevail. This happens for both pusher- and puller-type swimmers. For weak confinement, these excursions are symmetric, while they become asymmetric at stronger confinement, whereby the swimmer is located closer to one of the two walls. In this study, we combine  numerical and theoretical analyses.
\end{abstract}

\pacs{47.63.mh, 47.63.Gd, 47.15.G--,47.63.mf}

\maketitle


Some unicellular micro-organisms move on solid surfaces or swim in liquids by deforming their body instead of using flagella or cilia--this is known as {\it amoeboid motion}. Algae such as {\it Eutreptiella Gymnastica} \cite{Throndsen1969},
amoebae such as {\it dictyostelium discoideum} \cite{Barry2010,Baea2010}, but also leucocytes \cite{Barry2010,Baea2010} and even cancer cells \cite{Pinner2008}
use this specific way of locomotion. This is a complex movement that recently incited several theoretical studies
 \cite{Ohta2009,Hiraiwa2011,Shapere1987,Avron2004,Alouges2011,Loheac2013,Vilfan2012,Farutin2013} since it is intimately linked to cell migration involved in several diseases. Some experimental
results indicate that adhesion to a solid substratum is not a prerequisite for cells such as amoebae \cite{Barry2010} to produce
an amoeboid movement during cell migration and suggest that crawling close to a surface and
swimming are similar processes. Recently, it was shown that integrin (a protein involved in adhesion process) should no
longer be viewed as force transducers during locomotion but as
switchable immobilizing anchors that  slow down cells in the blood stream before transmigration. Indeed, leukocytes migrate by swimming in the absence of specific adhesive
interactions with the extracellular environment \cite{Lammermann2008}.

When moving, all micro-organisms are sensitive to their environments. Most microswimmers can follow gradients of chemicals (chemotaxis), some microalgae can move toward light sources (phototaxis) \cite{garcia2013} or orient themselves in the gravity field (gravitaxis) \cite{Kessler1985}, some other bacteria move along adhesion gradients (haptotaxis) \cite{McCarthy1983,Cantat1999}, etc. Spatial confinement
is another major environmental constraint which strongly influences the motion of micro-organisms. As a matter of fact, in the low-Reynolds number world, amoeboid motion generally occurs close to surfaces, in small capillaries or in extracellular matrices of biological tissues. Micro-organisms swim through permeable boundaries, cell walls or micro-vasculature. Therefore, the effect of walls on motile microorganisms has been a topic of increasingly active research \cite{Felderhof2010, Jana2012, Zottl2012, Zhu2013, Bilbao2013, Ledesma2013, Acemoglu2014,Liu2014,Zottl2014,Temel2015}. It has been calculated a long time ago by Katz \cite{Katz1974} and more recently pointed out \cite {Lauga2009,Ledesma2013,Zhu2013,Liu2014,Temel2015} that swimmers can take advantage of walls to increase their motility. Understanding the behavior of microswimmers in confinement can also pave the way to novel applications in microfluidic devices where properly shaped microstructures can interfere with swimming bacteria and guide, concentrate, and arrange populations of cells \cite{Wan2008}. Living microswimmers show a large variety of swimming strategies \cite{Lauga2009} as do theoretical models aiming at describing their  dynamics in confinement.

Felderhof \cite{Felderhof2010} has shown that the speed of Taylor-like swimmer increases with confinement. Zhu {\it{et al.}} \cite{Zhu2013} used the squirmer model to show that (when only tangential surface motion is included) the velocity decreases with confinement and that a pusher crashes into the wall, a puller settles in a straight trajectory, and a neutral swimmer navigates. When including normal deformation they found an increase of velocity with confinement. Liu {\it{et al.}} \cite{Liu2014} analyzed a helical flagellum in a tube and  found that except for a small range of tube radii,
the swimming speed, when the helix rotation rate is fixed, increases monotonically as the confinement
becomes tighter. Acemoglu {\it{et al.}} \cite{Acemoglu2014} adopted  a similar model but, besides the flagellum, they included a head and found a {\it decrease} of velocity with confinement. Bilbao {\it{et al.}}\cite{Bilbao2013} treated numerically a model inspired by nematode locomotion and found that it navigates more efficiently and moves faster due to walls. Ledesma {\it{et al.}}\cite{Ledesma2013} reported on a dipolar swimmer in a rigid or elastic tube and found a speed enhancement due to walls.

Here, we investigate, by means of numerical and analytical modeling, the effect of confinement on the behavior of an amoeboid swimmer, which is a deformable object subjected to active forces along its inextensible membrane. Our model swimmer is found to reveal interesting features when confined between two walls. (i) We find that straight trajectories might be unstable, independently of the nature of the swimmer (pusher or puller). (ii) For weak confinement, the swimming speed can either increase or decrease depending on the confinement strength. For strongly confined regimes, the velocity decreases in all cases recalling previous results on different models. (iii) The confined environment is shown to induce a transition from one to another type of swimmer (i.e. puller or pusher). These behaviors are unique to amoeboid swimming (AS) and point to a nontrivial dynamics owing to the internal degrees of freedom that evolve in response to various constraints.

\paragraph{The model.} Amoeboid swimming is modeled here by taking a one-dimensional ($1$D) closed and inextensible membrane, which encloses a two-dimensional ($2$D) liquid of certain viscosity $\eta$ and is suspended in another fluid taken to be of the same viscosity, for simplicity. The extra computational complexity of dealing with confined geometry restricts our study to $2$D, which draws already rich behaviors. The effective radius of the swimmer is $R_0=\sqrt{A_0/\pi}$, where $A_0$ is the enclosed area. The swimmer has an excess normalized perimeter $\Gamma=L_0/(2\pi R_0)-1$ ($L_0$ is the perimeter) with respect to a circular shape ($\Gamma=0$ corresponds to a circle, whereas large $\Gamma$ signifies a very deflated, and thus amply deformable, swimmer). The strength of confinement is defined as $C=2R_0/W$, with $W$ the channel width. Bounding walls are parallel to the $x$ direction and $y$ denotes the orthogonal one.

A set of active forces is distributed on the membrane that reacts with tension forces to preserve the local arclength.
The total force density is given by
\begin{equation}
{\mathbf F}=F_a {\mathbf n} - \zeta c {\mathbf n} + \frac{\partial \zeta} { \partial s}{\mathbf t},
\end{equation}
where ${ F_a} {\mathbf n}$ is the active force to be specified below (which we take to point along the normal ${\mathbf n}$ for simplicity), $\zeta$ is a Lagrange multiplier that enforces local membrane incompressibility, $c$ is the curvature, ${\mathbf t}$ is the unit tangent vector and $s$ is the arclength. We impose zero  total force  and torque. In its full generality the active force can be decomposed into a Fourier series ${ F}_a(\alpha,t)= \sum _{k=-k_{max}} ^{k=k_{max}}{ F}_k(t) e^{ik\alpha}$ with $\alpha =2\pi s/L_0$. We first consider the case $k_{max}=3$, so that we are left with two complex amplitudes  $F_2$ and $F_3$. Other configurations of the forces have been explored as well (see below). We consider cyclic strokes represented by $F_2= F_{-2}=-A\cos (\omega t)$ and $F_3=F_{-3}= A\sin (\omega t)$, where $A$ is the force amplitude.

\begin{figure}
\begin{center}
\includegraphics[angle=0,width=0.9\columnwidth]{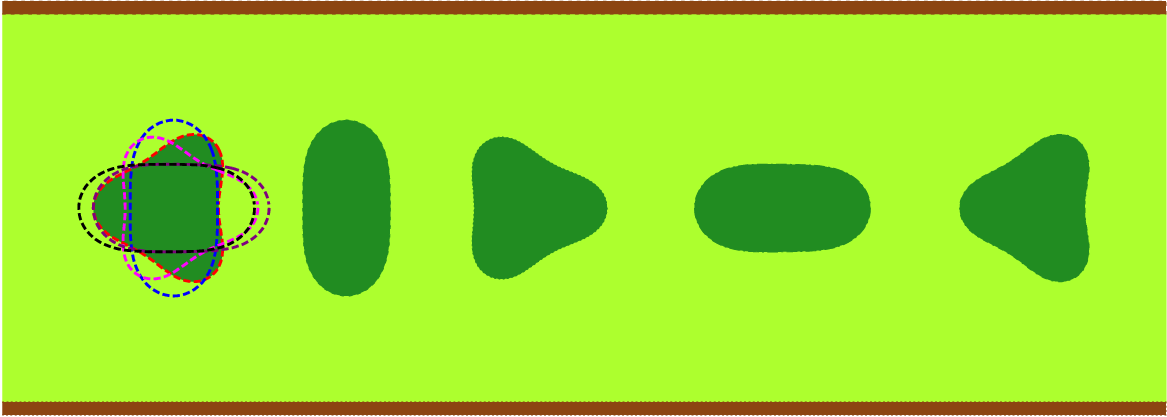}
\caption{\label{snap} (Color online) Snapshots of an axially moving swimmer over time ($W=6R_{0}$). The dashed profiles show a complete period $T_s$ of deformation and then a few shapes are represented over a time of the order of $75 T_{s}$. $\Gamma=0.085$.}
\end{center}
\end{figure}

\begin{figure}[b]
\begin{center}
\includegraphics[angle=0,width=0.9\columnwidth]{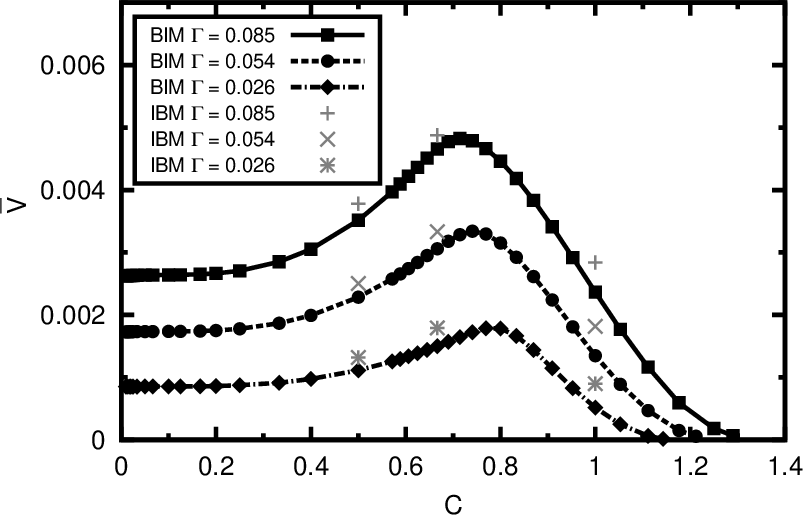}
\caption{\label{v_central} Time-averaged velocity magnitudes (as a function of confinement $C$) of an axially moving swimmer for different $\Gamma$ values.
}
\end{center}
\end{figure}

The Stokes equations with boundary conditions (force balance condition, continuity of the fluid velocity and membrane incompressibility) are solved  using either  the boundary integral method (BIM)~\cite{Thiebaud2013}
or the immersed boundary method (IBM) \cite{Lai2010}.

Besides $\Gamma$ and $C$, there is an additional dimensionless number $S=A/(\omega \eta)$, which is the ratio between the time scale associated with swimming strokes ($T_s= 2\pi /\omega$) and the time scale of fluid flow due to active force ($T_c=\eta/A$). Here we take $S=10.0$ (the shape has enough time to respond to active forces) and explore the effects of $\Gamma$ and $C$. At a large distance from the swimmer, the velocity field is governed by  $\sigma_{ij}=\oint F_ir_j ds$.  Only the (dimensionless) stresslet $\Sigma=(\sigma_{xx}-\sigma_{yy})/(\eta /T_s)$ enters the velocity field for symmetric swimmers. ${\Sigma}>0$ defines a pusher and ${ \Sigma}<0$ defines a puller. The force distribution defined above is found to correspond to a pusher in the absence of walls. Below we will see how to monitor a puller or pusher and how the walls change the nature of the swimmer.

\begin{figure}
\begin{center}
\includegraphics[angle=0,width=1\columnwidth]{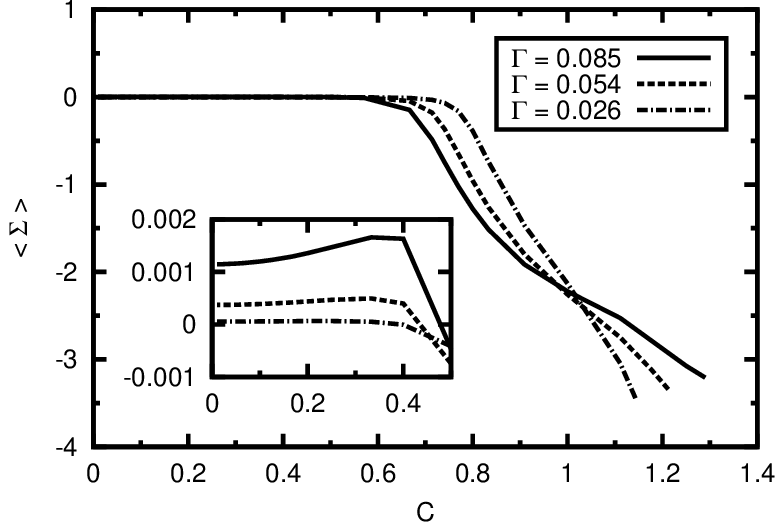}
\caption{\label{stresslet} Time-averaged $\langle\Sigma\rangle$ as a function of confinement $C$ showing the transition from pusher to puller.}
\end{center}
\end{figure}

\begin{figure}[b]
\begin{center}
\includegraphics[angle=0,width=0.9\columnwidth]{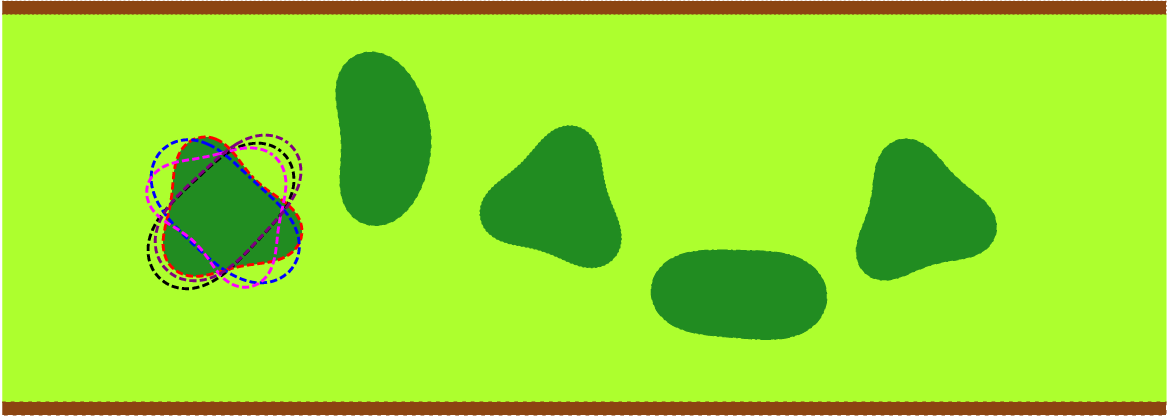}
\caption{\label{snap2} (Color online) Snapshots of a navigating swimmer over time ($W=6R_{0}$). The dashed profiles show a complete period $T_s$ of deformation and then a few shapes are represented over a navigation period $T$ of the order of $50 T_s $. $\Gamma=0.085$.}
\end{center}
\end{figure}

\paragraph{Results: Axially moving swimmers.} We first consider an axially moving swimmer (AMS) (see Fig.~\ref{snap}). We consider only dimensionless quantities (unless otherwise stated). For example, $\bar{V}=V T_c/R_0$ will denote the magnitude of swimming speed. We find  an optimal confinement for swimming velocity. Increasing  $C$ enhances the speed of the swimmer until an optimal $C^{o}$, where the speed attains a maximum before it decreases. Around the optimal value  $C^{o}$, low (high) viscous friction between the swimmer and the walls during the forward (recovery) phase of swimming promotes AMS speed. When the confinement is too strong, large amplitude deformations are frustrated  resulting in a loss of speed. The velocity collapse at strong confinement was also reported for helical flagellum \cite{Liu2014,Acemoglu2014} and is expected to occur for all swimmer models. Figure \ref{v_central} shows the swimming velocity magnitude  for different $\Gamma$ values. That the wall enhances motility seems to be a quite  general fact, as  reported in the literature \cite{Felderhof2010,Jana2012,Zhu2013,Bilbao2013,Ledesma2013,Acemoglu2014,Liu2014}.  However, we must stress that this is not a systematic tendency. Close inspection shows that at weak confinement velocity first decreases before increasing, as shown in Ref.~\cite{sm}.

\begin{figure}
\begin{center}
\includegraphics[angle=0,width=0.9\columnwidth]{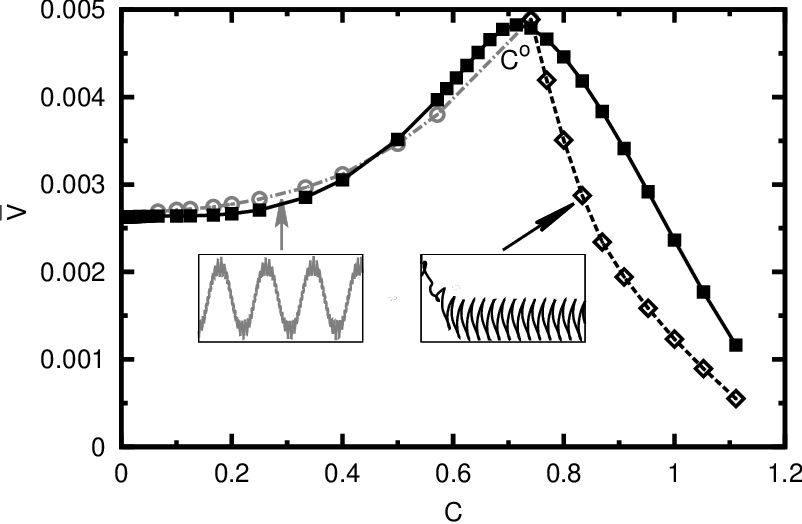}
\caption{\label{zigzag} Time-averaged velocity magnitudes (as a function of confinement $C$) of different swimmers ($\Gamma=0.085$): migrating along one wall (black diamond dashed line), navigating between two walls (gray circle dotted dashed line) and moving along the channel center (black square solid line). The insets show characteristic trajectories.}
\end{center}
\end{figure}

\begin{figure}[b]
\begin{center}
\includegraphics[angle=0,width=0.9\columnwidth]{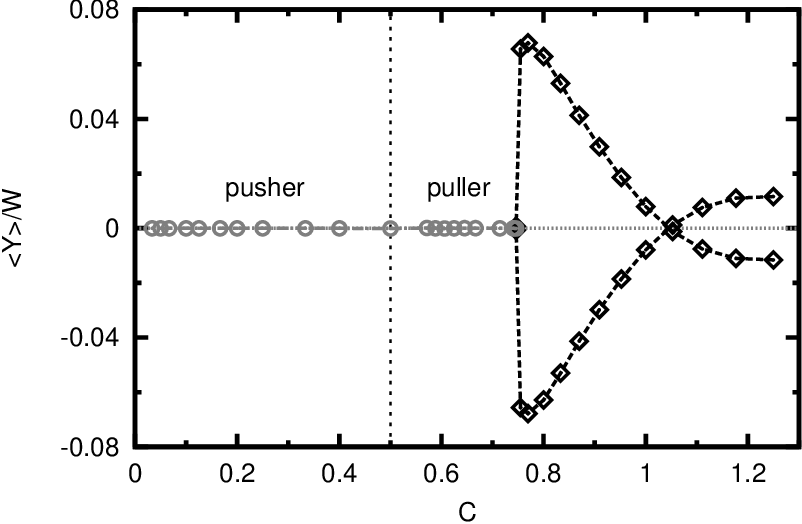}
\caption{\label{bifurcation} Average position of the center of mass over a navigation period as a function of confinement $C$. $\Gamma=0.085$. Circles (diamonds) correspond to the symmetric (asymmetric) motion of the swimmer. The vertical dotted line is the demarcation line between a pusher and puller. Note that a puller can either navigate or move close to either wall.}
\end{center}
\end{figure}

\paragraph{Results: Swimmer nature evolution.} The value of the dimensionless stresslet  $\Sigma$ depends on the instantaneous swimmer configuration and its sign instructs us on the nature of the swimmer. We determine the average stresslet over a navigation cycle. An interesting result is the effect of confinement on the pusher/puller nature of the swimmer. For small $C (<0.5)$ the swimmer is found to behave as a pusher, while  it behaves as a puller for larger $C (>0.5)$. Figure~\ref{stresslet} shows the evolution of $\langle\Sigma\rangle$ as a function of confinement, where  a transition from pusher to puller is observed. 

\paragraph{Results: Instability of the central position.} The central position after a long time is found to be unstable.  The swimmer exhibits at small $C$ (weakly confined regime) a zigzag motion undergoing large amplitude excursions from one wall to the other. We refer to this as a {\it navigating swimmer} (NS). Figure~\ref{snap2} shows a snapshot, whereas the insets of Fig.~\ref{zigzag} display typical trajectories. Despite this complex motion, the velocity in Fig. \ref{zigzag} behaves with $C$ {qualitatively} as that of the central swimmer.

The NS trajectory was recently reported \cite{Najafi2013,Zhu2013}  { in the cases of squirmer and three-bead models and also observed experimentally for paramecium (ciliated motility) in a tube \cite{Jana2012} pointing to the genericity of navigation}.  { This instability can be explained analytically (see Ref.~\cite{sm})}. 

A remarkable property is  that the navigation mode can be adopted both by the pusher and the puller. This is in contrast with nonamoeboid motion \cite{Zhu2013}, where a pusher is found to crash into the wall whereas the puller settles into a straight trajectory. These last two behaviors are also recovered by our simulations, provided the stresslet amplitude is large enough (${\Sigma ^2}>>-\bar{V}DS$, where  $D$ is the dimensionless force quadrupole strength; see Ref.~\cite{sm}).

\paragraph{Results: Symmetry-breaking bifurcation.} At a critical {$C^{*}$} the symmetric excursion of the swimmer becomes unstable and undergoes a bifurcation characterized by the loss of the central symmetry in favor of an asymmetric excursion in the channel, as shown in  the trajectories of Fig.~\ref{zigzag} (see the supplemental movies in Ref.~\cite{sm}). Figure \ref{bifurcation} shows the average position in $y$ of the center of mass as a function of confinement: a bifurcation diagram. This bifurcation is very abrupt, albeit it is of supercritical nature. Both slightly before and beyond the bifurcation, the swimmer behaves on average as a puller, but still it exhibits two very distinct modes of locomotion: navigation or settling into a quasistraight trajectory (oscillation of the center of mass in this regime is fixed by the amoeboid cycle). This complexity is triggered by the intricate nature of the amoeboid degrees of freedom.


\begin{figure}
\hspace{-2cm}
\includegraphics[angle=0,width=0.7\columnwidth]{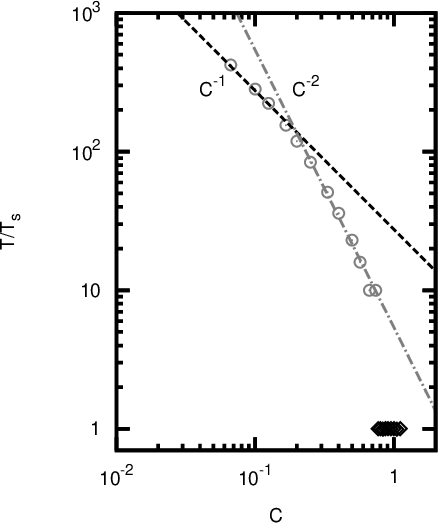}
\caption{\label{period} Period of navigation as a function of confinement $C$. $\Gamma=0.085$. Circles (diamonds) correspond to the symmetric (asymmetric) motion of the swimmer.}
\end{figure}

\paragraph{Results: Other force distributions.}
Including force distributions up to sixth harmonics with various amplitudes leaves the overall picture  unchanged, pointing to the generic character of AS. The next step has consisted in linking the nature of the swimmer to its dynamics. We have monitored  a pusher or puller type of swimmer. If ${\mathbf F} =2 [ \sin(\omega t) \cos (3\alpha)-(\beta + \cos (\omega t )) \cos(2\alpha)] {\mathbf n}$, we have a puller; while if ${\mathbf F }=2[ (-\beta + \sin(\omega t)) \cos (3\alpha)- \cos (\omega t ) \cos(2\alpha)] {\mathbf n}$, we have a pusher (with $\beta >0$). $\beta$ monitors the strength of the swimmer nature (weak and strong  pusher  or puller). We found that for a weak enough stresslet, symmetric and asymmetric navigations prevail both for pullers and pushers. For a strong enough stresslet amplitude (for $\beta>\beta_c\sim 1$), we find that the pusher crashes into the wall, while a  puller settles into a straight trajectory. This means that there is a qualitative change of behavior triggered by $\beta$, on which  we shall report on systematic study in the future.

\paragraph{Results: Navigation period.}
The navigation period $T$ exhibits a nontrivial behavior with $C$ (Fig. \ref{period}).  At small $C$, the period scales as $T\sim C^{-1}$, and as  $T\sim C^{-2}$ at intermediate confinement, before attaining a plateau at stronger confinement. To dig into the reasons for this complex behavior, we provide here some heuristic arguments. In the first regime, the NS swims in a straight and monotonous manner towards the next wall. In that regime, the period is limited by the distance traveled by the swimmer of the order $ W= 2R/ C $. This naturally yields the $C^{-1}$ scaling of Fig. \ref{period} for weak $C$. In the intermediate confinement regime, the magnitude of velocity depends linearly on $C$, so that the period scales as $C^{-2}$ (see also Ref.~\cite{sm}). After the symmetry-breaking occurs, the NS stays close to one of the two walls (inset of Fig.~\ref{zigzag}), and its center of mass oscillates with the intrinsic stroke period $T_{s}$. In this regime, the  period is independent of $C$ (diamonds in Fig.~\ref{period}).

\paragraph{Analytical results.} We have first performed a linear stability analysis \cite{sm}. We find, for small $C$, that the stability of the swimmer is governed by the stresslet sign: For ${\Sigma}>0$ (pusher) the straight trajectory is unstable, while it is stable otherwise (puller) . For a neutral swimmer the trajectory is marginally stable (the stability eigenvalue $\Omega$, for a  perturbation of the form $y \sim e^{\Omega t}$, is  purely imaginary). We find that in the intermediate $C$ regime the navigation period  behaves as $C^{-2}.$ Using a systematic multipole expansion the complex behavior of the velocity as a function of $C$ (at low $C$) can be explained \cite{sm}.


\paragraph{Discussion.} We believe that the global features revealed by our study will persist in $3$D although extending our work to $3$D simulations will be a challenging task. Besides, in order to better match real cells performing amoeboid swimming (e.g. leukocytes), cytoskeleton dynamics and its relation to force generation will be important ingredients to be included in a $3$D modeling.


\begin{acknowledgments}
C.M., A.F., M.T. and H.W. are grateful to CNES and ESA for financial support. S.R. and P.P. acknowledge financial support from ANR. All the authors acknowledge the French-Taiwanese  ORCHID cooperation grant. W.-F. H., M.-C.L. and C.M. thank the MoST for a financial support allowing initiation of this project.
\end{acknowledgments}

\bibliography{chlamybib}

\begin{thebibliography}{34}
\expandafter\ifx\csname natexlab\endcsname\relax\def\natexlab#1{#1}\fi
\expandafter\ifx\csname bibnamefont\endcsname\relax
  \def\bibnamefont#1{#1}\fi
\expandafter\ifx\csname bibfnamefont\endcsname\relax
  \def\bibfnamefont#1{#1}\fi
\expandafter\ifx\csname citenamefont\endcsname\relax
  \def\citenamefont#1{#1}\fi
\expandafter\ifx\csname url\endcsname\relax
  \def\url#1{\texttt{#1}}\fi
\expandafter\ifx\csname urlprefix\endcsname\relax\def\urlprefix{URL }\fi
\providecommand{\bibinfo}[2]{#2}
\providecommand{\eprint}[2][]{\url{#2}}

\bibitem[{\citenamefont{Throndsen}(1969)}]{Throndsen1969}
\bibinfo{author}{\bibfnamefont{J.}~\bibnamefont{Throndsen}},
  \bibinfo{journal}{Norw. J. Bot.} \textbf{\bibinfo{volume}{16}},
  \bibinfo{pages}{161} (\bibinfo{year}{1969}).

\bibitem[{\citenamefont{Barry and Bretscher}(2010)}]{Barry2010}
\bibinfo{author}{\bibfnamefont{N.~P.} \bibnamefont{Barry}} \bibnamefont{and}
  \bibinfo{author}{\bibfnamefont{M.~S.} \bibnamefont{Bretscher}},
  \bibinfo{journal}{Proc. Natl. Acad. Sci. U.S.A.}
  \textbf{\bibinfo{volume}{107}}, \bibinfo{pages}{11 376}
  (\bibinfo{year}{2010}).

\bibitem[{\citenamefont{Baea and Bodenschatz}(2010)}]{Baea2010}
\bibinfo{author}{\bibfnamefont{A.~J.} \bibnamefont{Baea}} \bibnamefont{and}
  \bibinfo{author}{\bibfnamefont{E.}~\bibnamefont{Bodenschatz}},
  \bibinfo{journal}{Proc. Natl. Acad. Sci. U.S.A.}
  \textbf{\bibinfo{volume}{107}}, \bibinfo{pages}{E167} (\bibinfo{year}{2010}).

\bibitem[{\citenamefont{Pinner and Sahai}(2008)}]{Pinner2008}
\bibinfo{author}{\bibfnamefont{S.}~\bibnamefont{Pinner}} \bibnamefont{and}
  \bibinfo{author}{\bibfnamefont{E.}~\bibnamefont{Sahai}}, \bibinfo{journal}{J.
  Micros.} \textbf{\bibinfo{volume}{231}}, \bibinfo{pages}{441}
  (\bibinfo{year}{2008}).

\bibitem[{\citenamefont{Ohta and Ohkuma}(2009)}]{Ohta2009}
\bibinfo{author}{\bibfnamefont{T.}~\bibnamefont{Ohta}} \bibnamefont{and}
  \bibinfo{author}{\bibfnamefont{T.}~\bibnamefont{Ohkuma}},
  \bibinfo{journal}{Phys. Rev. Lett.} \textbf{\bibinfo{volume}{102}},
  \bibinfo{pages}{154101} (\bibinfo{year}{2009}).

\bibitem[{\citenamefont{Hiraiwa et~al.}(2011)\citenamefont{Hiraiwa, Shitaraa,
  and Ohta}}]{Hiraiwa2011}
\bibinfo{author}{\bibfnamefont{T.}~\bibnamefont{Hiraiwa}},
  \bibinfo{author}{\bibfnamefont{K.}~\bibnamefont{Shitaraa}}, \bibnamefont{and}
  \bibinfo{author}{\bibfnamefont{T.}~\bibnamefont{Ohta}},
  \bibinfo{journal}{Soft Matter} \textbf{\bibinfo{volume}{7}},
  \bibinfo{pages}{3083} (\bibinfo{year}{2011}).

\bibitem[{\citenamefont{Shapere and Wilczek}(1987)}]{Shapere1987}
\bibinfo{author}{\bibfnamefont{A.}~\bibnamefont{Shapere}} \bibnamefont{and}
  \bibinfo{author}{\bibfnamefont{F.}~\bibnamefont{Wilczek}},
  \bibinfo{journal}{Phys. Rev. Lett.} \textbf{\bibinfo{volume}{58}},
  \bibinfo{pages}{2051} (\bibinfo{year}{1987}).

\bibitem[{\citenamefont{Avron et~al.}(2004)\citenamefont{Avron, Gat, and
  Kenneth}}]{Avron2004}
\bibinfo{author}{\bibfnamefont{J.~E.} \bibnamefont{Avron}},
  \bibinfo{author}{\bibfnamefont{O.}~\bibnamefont{Gat}}, \bibnamefont{and}
  \bibinfo{author}{\bibfnamefont{O.}~\bibnamefont{Kenneth}},
  \bibinfo{journal}{Phys. Rev. Lett.} \textbf{\bibinfo{volume}{93}},
  \bibinfo{pages}{186001} (\bibinfo{year}{2004}).

\bibitem[{\citenamefont{Alouges et~al.}(2011)\citenamefont{Alouges, Desimone,
  and Heltai}}]{Alouges2011}
\bibinfo{author}{\bibfnamefont{F.}~\bibnamefont{Alouges}},
  \bibinfo{author}{\bibfnamefont{A.}~\bibnamefont{Desimone}}, \bibnamefont{and}
  \bibinfo{author}{\bibfnamefont{L.}~\bibnamefont{Heltai}},
  \bibinfo{journal}{Math. Models Methods Appl. Sci.}
  \textbf{\bibinfo{volume}{21}}, \bibinfo{pages}{361} (\bibinfo{year}{2011}).

\bibitem[{\citenamefont{Loheac et~al.}(2013)\citenamefont{Loheac, Scheid, and
  Tucsnak}}]{Loheac2013}
\bibinfo{author}{\bibfnamefont{J.}~\bibnamefont{Loheac}},
  \bibinfo{author}{\bibfnamefont{J.-F.} \bibnamefont{Scheid}},
  \bibnamefont{and} \bibinfo{author}{\bibfnamefont{M.}~\bibnamefont{Tucsnak}},
  \bibinfo{journal}{Acta Appl. Math.} \textbf{\bibinfo{volume}{123}},
  \bibinfo{pages}{175} (\bibinfo{year}{2013}).

\bibitem[{\citenamefont{Vilfan}(2012)}]{Vilfan2012}
\bibinfo{author}{\bibfnamefont{A.}~\bibnamefont{Vilfan}},
  \bibinfo{journal}{Phys. Rev. Lett.} \textbf{\bibinfo{volume}{109}},
  \bibinfo{pages}{128105} (\bibinfo{year}{2012}).

\bibitem[{\citenamefont{Farutin et~al.}(2013)\citenamefont{Farutin, Rafai,
  Dysthe, Duperray, Peyla, and Misbah}}]{Farutin2013}
\bibinfo{author}{\bibfnamefont{A.}~\bibnamefont{Farutin}},
  \bibinfo{author}{\bibfnamefont{S.}~\bibnamefont{Rafai}},
  \bibinfo{author}{\bibfnamefont{D.~K.} \bibnamefont{Dysthe}},
  \bibinfo{author}{\bibfnamefont{A.}~\bibnamefont{Duperray}},
  \bibinfo{author}{\bibfnamefont{P.}~\bibnamefont{Peyla}}, \bibnamefont{and}
  \bibinfo{author}{\bibfnamefont{C.}~\bibnamefont{Misbah}},
  \bibinfo{journal}{Phys. Rev. Lett.} \textbf{\bibinfo{volume}{111}},
  \bibinfo{pages}{228102} (\bibinfo{year}{2013}).

\bibitem[{\citenamefont{L\"ammermann et~al.}(2008)\citenamefont{L\"ammermann,
  Bader, Monkley, Worbs, Wedlich-Soldner, Hirsch, Keller, Forster, Critchley,
  Fassler et~al.}}]{Lammermann2008}
\bibinfo{author}{\bibfnamefont{T.}~\bibnamefont{L\"ammermann}},
  \bibinfo{author}{\bibfnamefont{B.~L.} \bibnamefont{Bader}},
  \bibinfo{author}{\bibfnamefont{S.~J.} \bibnamefont{Monkley}},
  \bibinfo{author}{\bibfnamefont{T.}~\bibnamefont{Worbs}},
  \bibinfo{author}{\bibfnamefont{R.}~\bibnamefont{Wedlich-Soldner}},
  \bibinfo{author}{\bibfnamefont{K.}~\bibnamefont{Hirsch}},
  \bibinfo{author}{\bibfnamefont{M.}~\bibnamefont{Keller}},
  \bibinfo{author}{\bibfnamefont{R.}~\bibnamefont{Forster}},
  \bibinfo{author}{\bibfnamefont{D.~R.} \bibnamefont{Critchley}},
  \bibinfo{author}{\bibfnamefont{R.}~\bibnamefont{Fassler}},
  \bibnamefont{et~al.}, \bibinfo{journal}{Nature (London)}
  \textbf{\bibinfo{volume}{453}}, \bibinfo{pages}{51} (\bibinfo{year}{2008}).

\bibitem[{\citenamefont{Garcia et~al.}(2013)\citenamefont{Garcia, Rafa\"i, and
  Peyla}}]{garcia2013}
\bibinfo{author}{\bibfnamefont{X.}~\bibnamefont{Garcia}},
  \bibinfo{author}{\bibfnamefont{S.}~\bibnamefont{Rafa\"i}}, \bibnamefont{and}
  \bibinfo{author}{\bibfnamefont{P.}~\bibnamefont{Peyla}},
  \bibinfo{journal}{Phys. Rev. Lett.} \textbf{\bibinfo{volume}{110}},
  \bibinfo{pages}{138106} (\bibinfo{year}{2013}).

\bibitem[{\citenamefont{Kessler}(1985)}]{Kessler1985}
\bibinfo{author}{\bibfnamefont{J.}~\bibnamefont{Kessler}},
  \bibinfo{journal}{Contemp. Phys.} \textbf{\bibinfo{volume}{26}},
  \bibinfo{pages}{147} (\bibinfo{year}{1985}).

\bibitem[{\citenamefont{McCarthy et~al.}(1983)\citenamefont{McCarthy, Palm, and
  Furcht}}]{McCarthy1983}
\bibinfo{author}{\bibfnamefont{J.~B.} \bibnamefont{McCarthy}},
  \bibinfo{author}{\bibfnamefont{S.~L.} \bibnamefont{Palm}}, \bibnamefont{and}
  \bibinfo{author}{\bibfnamefont{L.~T.} \bibnamefont{Furcht}},
  \bibinfo{journal}{J. Cell Biol.} \textbf{\bibinfo{volume}{97}},
  \bibinfo{pages}{772} (\bibinfo{year}{1983}).

\bibitem[{\citenamefont{Cantat and Misbah}(1999)}]{Cantat1999}
\bibinfo{author}{\bibfnamefont{I.}~\bibnamefont{Cantat}} \bibnamefont{and}
  \bibinfo{author}{\bibfnamefont{C.}~\bibnamefont{Misbah}},
  \bibinfo{journal}{Phys. Rev. Lett.} \textbf{\bibinfo{volume}{83}},
  \bibinfo{pages}{235} (\bibinfo{year}{1999}).

\bibitem[{\citenamefont{Felderhof}(2010)}]{Felderhof2010}
\bibinfo{author}{\bibfnamefont{B.~U.} \bibnamefont{Felderhof}},
  \bibinfo{journal}{Phys. Fluids} \textbf{\bibinfo{volume}{22}},
  \bibinfo{eid}{113604} (\bibinfo{year}{2010}).

\bibitem[{\citenamefont{Jana et~al.}(2012)\citenamefont{Jana, Um, and
  Jung}}]{Jana2012}
\bibinfo{author}{\bibfnamefont{S.}~\bibnamefont{Jana}},
  \bibinfo{author}{\bibfnamefont{S.~H.} \bibnamefont{Um}}, \bibnamefont{and}
  \bibinfo{author}{\bibfnamefont{S.}~\bibnamefont{Jung}},
  \bibinfo{journal}{Phys. Fluids} \textbf{\bibinfo{volume}{24}},
  \bibinfo{eid}{041901} (\bibinfo{year}{2012}).

\bibitem[{\citenamefont{Z\"ottl and Stark}(2012)}]{Zottl2012}
\bibinfo{author}{\bibfnamefont{A.}~\bibnamefont{Z\"ottl}} \bibnamefont{and}
  \bibinfo{author}{\bibfnamefont{H.}~\bibnamefont{Stark}},
  \bibinfo{journal}{Phys. Rev. Lett.} \textbf{\bibinfo{volume}{108}},
  \bibinfo{pages}{218104} (\bibinfo{year}{2012}).

\bibitem[{\citenamefont{Zhu et~al.}(2013)\citenamefont{Zhu, Lauga, and
  Brandt}}]{Zhu2013}
\bibinfo{author}{\bibfnamefont{L.}~\bibnamefont{Zhu}},
  \bibinfo{author}{\bibfnamefont{E.}~\bibnamefont{Lauga}}, \bibnamefont{and}
  \bibinfo{author}{\bibfnamefont{L.}~\bibnamefont{Brandt}},
  \bibinfo{journal}{J. Fluid Mech.} \textbf{\bibinfo{volume}{726}},
  \bibinfo{eid}{011701} (\bibinfo{year}{2013}).

\bibitem[{\citenamefont{Bilbao et~al.}(2013)\citenamefont{Bilbao, Wajnryb,
  Vanapalli, and Blawzdziewicz}}]{Bilbao2013}
\bibinfo{author}{\bibfnamefont{A.}~\bibnamefont{Bilbao}},
  \bibinfo{author}{\bibfnamefont{E.}~\bibnamefont{Wajnryb}},
  \bibinfo{author}{\bibfnamefont{S.~A.} \bibnamefont{Vanapalli}},
  \bibnamefont{and}
  \bibinfo{author}{\bibfnamefont{J.}~\bibnamefont{Blawzdziewicz}},
  \bibinfo{journal}{Phys. Fluids} \textbf{\bibinfo{volume}{25}},
  \bibinfo{eid}{081902} (\bibinfo{year}{2013}).

\bibitem[{\citenamefont{Ledesma-Aguilar and Yeomans}(2013)}]{Ledesma2013}
\bibinfo{author}{\bibfnamefont{R.}~\bibnamefont{Ledesma-Aguilar}}
  \bibnamefont{and} \bibinfo{author}{\bibfnamefont{J.~M.}
  \bibnamefont{Yeomans}}, \bibinfo{journal}{Phys. Rev. Lett.}
  \textbf{\bibinfo{volume}{111}}, \bibinfo{pages}{138101}
  (\bibinfo{year}{2013}).

\bibitem[{\citenamefont{Acemoglu and Yesilyurt}(2014)}]{Acemoglu2014}
\bibinfo{author}{\bibfnamefont{A.}~\bibnamefont{Acemoglu}} \bibnamefont{and}
  \bibinfo{author}{\bibfnamefont{S.}~\bibnamefont{Yesilyurt}},
  \bibinfo{journal}{Biophys. J.} \textbf{\bibinfo{volume}{106}},
  \bibinfo{pages}{1537} (\bibinfo{year}{2014}).

\bibitem[{\citenamefont{Liu et~al.}(2014)\citenamefont{Liu, Breuer, and
  Powers}}]{Liu2014}
\bibinfo{author}{\bibfnamefont{B.}~\bibnamefont{Liu}},
  \bibinfo{author}{\bibfnamefont{K.~S.} \bibnamefont{Breuer}},
  \bibnamefont{and} \bibinfo{author}{\bibfnamefont{T.~R.}
  \bibnamefont{Powers}}, \bibinfo{journal}{Phys. Fluids}
  \textbf{\bibinfo{volume}{26}}, \bibinfo{eid}{011701} (\bibinfo{year}{2014}).

\bibitem[{\citenamefont{Z\"ottl and Stark}(2014)}]{Zottl2014}
\bibinfo{author}{\bibfnamefont{A.}~\bibnamefont{Z\"ottl}} \bibnamefont{and}
  \bibinfo{author}{\bibfnamefont{H.}~\bibnamefont{Stark}},
  \bibinfo{journal}{Phys. Rev. Lett.} \textbf{\bibinfo{volume}{112}},
  \bibinfo{pages}{118101} (\bibinfo{year}{2014}).

\bibitem[{\citenamefont{Temel and Yesilyurt}(2015)}]{Temel2015}
\bibinfo{author}{\bibfnamefont{F.~Z.} \bibnamefont{Temel}} \bibnamefont{and}
  \bibinfo{author}{\bibfnamefont{S.}~\bibnamefont{Yesilyurt}},
  \bibinfo{journal}{Bioinspiration Biomimetics} \textbf{\bibinfo{volume}{726}},
  \bibinfo{pages}{016015} (\bibinfo{year}{2015}).

\bibitem[{\citenamefont{Katz}(2013)}]{Katz1974}
\bibinfo{author}{\bibfnamefont{D.}~\bibnamefont{Katz}}, \bibinfo{journal}{J .
  Fluid Mech} \textbf{\bibinfo{volume}{64}}, \bibinfo{pages}{33}
  (\bibinfo{year}{2013}).

\bibitem[{\citenamefont{Lauga and Powers}(2009)}]{Lauga2009}
\bibinfo{author}{\bibfnamefont{E.}~\bibnamefont{Lauga}} \bibnamefont{and}
  \bibinfo{author}{\bibfnamefont{T.}~\bibnamefont{Powers}},
  \bibinfo{journal}{Rep. Prog. Phys.} \textbf{\bibinfo{volume}{72}},
  \bibinfo{pages}{096601} (\bibinfo{year}{2009}).

\bibitem[{\citenamefont{Wan et~al.}(2008)\citenamefont{Wan, Reichhardt,
  Nussinov, and Reichhardt}}]{Wan2008}
\bibinfo{author}{\bibfnamefont{M.~B.} \bibnamefont{Wan}},
  \bibinfo{author}{\bibfnamefont{C.~J.~O.} \bibnamefont{Reichhardt}},
  \bibinfo{author}{\bibfnamefont{Z.}~\bibnamefont{Nussinov}}, \bibnamefont{and}
  \bibinfo{author}{\bibfnamefont{C.}~\bibnamefont{Reichhardt}},
  \bibinfo{journal}{Phys. Rev. Lett.} \textbf{\bibinfo{volume}{101}},
  \bibinfo{pages}{018102} (\bibinfo{year}{2008}).

\bibitem[{\citenamefont{Thi\'ebaud and Misbah}(2013)}]{Thiebaud2013}
\bibinfo{author}{\bibfnamefont{M.}~\bibnamefont{Thi\'ebaud}} \bibnamefont{and}
  \bibinfo{author}{\bibfnamefont{C.}~\bibnamefont{Misbah}},
  \bibinfo{journal}{Phys. Rev. E} \textbf{\bibinfo{volume}{88}},
  \bibinfo{pages}{062707} (\bibinfo{year}{2013}).

\bibitem[{\citenamefont{Kim and Lai}(2010)}]{Lai2010}
\bibinfo{author}{\bibfnamefont{Y.}~\bibnamefont{Kim}} \bibnamefont{and}
  \bibinfo{author}{\bibfnamefont{M.-C.} \bibnamefont{Lai}},
  \bibinfo{journal}{J. Comp. Phys.} \textbf{\bibinfo{volume}{229}},
  \bibinfo{pages}{4840 } (\bibinfo{year}{2010}).

\bibitem[{sm()}]{sm}
\bibinfo{note}{See Supplemental Material at [URL will be inserted by publisher]
  for the theoretical analyses of confined swimming velocity and the relation
  between the stability of trajectory and the nature of swimmer shown in the
  paper as well as the movies for two different swimming modes.}

\bibitem[{\citenamefont{Najafi et~al.}(2013)\citenamefont{Najafi, Raad, and
  Yousefi}}]{Najafi2013}
\bibinfo{author}{\bibfnamefont{A.}~\bibnamefont{Najafi}},
  \bibinfo{author}{\bibfnamefont{S.~S.~H.} \bibnamefont{Raad}},
  \bibnamefont{and} \bibinfo{author}{\bibfnamefont{R.}~\bibnamefont{Yousefi}},
  \bibinfo{journal}{Phys. Rev. E} \textbf{\bibinfo{volume}{88}},
  \bibinfo{pages}{045001} (\bibinfo{year}{2013}).

\end{thebibliography}

\end{document}